
\let\rawfootnote=\footnote
\def\footnote#1#2{{\baselineskip=\normalbaselineskip
     \parindent=0pt\parskip=0pt\rm
     \rawfootnote{#1}{#2\hfill\vrule height 0pt depth 6pt width 0pt}}}
\magnification=\magstep1
\baselineskip=20pt
\centerline{Nova Dust Nucleation: Kinetics and Photodissociation}
\bigskip
\centerline{D. J. Johnson\footnote*{Department of Physics and McDonnell
Center for the Space Sciences, Washington University, St.~Louis, Mo.
63130}\footnote\dag{Present Address: Department of Natural Sciences,
St.~Thomas Aquinas College, Sparkill, N. Y. 10976}, M. W. Friedlander$^{\rm
*}$, and J. I. Katz$^{\rm *}$}
\bigskip
\centerline{Abstract}
\medskip
Dust is observed to form in nova ejecta.  The grain temperature is
determined by the diluted nova radiation field rather than the gas kinetic
temperature, making classical nucleation theory inapplicable.  We used
kinetic equations to calculate the growth of carbon nuclei in these ejecta.
For expected values of the parameters too many clusters grew, despite the
small sticking probability of atoms to small clusters, and the clusters only
reached radii of about 100\AA\ when the carbon vapor was depleted.  We then
included the effects of cluster photodissociation by ultraviolet radiation
from the nova.  This suppresses nucleation, but too well, and no grains form
at all.  Finally we suggest that a few growing carbon nuclei may be protected
from photodissociation by a sacrificial surface layer of hydrogen.
\vfil
\eject
\centerline{I. Introduction}
\medskip
Novae can be divided into two general classes: fast novae which fade more
rapidly than 0.1 magnitude day$^{-1}$ and have higher average luminosities,
and moderate speed and slow novae which fade less rapidly than 0.1 magnitude
day$^{-1}$ and have lower average luminosities.  The decline in the visual
light curve for moderate speed novae is accompanied by an increase in
infrared brightness, with the infrared luminosity matching the initial
bolometric luminosity of the nova.  These observations have been interpreted
as resulting from the condensation of dust grains in an expanding shell
of ejecta.  This generally accepted model is built on
the assumption that visible and ultraviolet radiation from the nova is
absorbed by the dust, and is reradiated in the infrared.  Fast novae, in
contrast, do not show this evidence for dust production.

Because infrared spectral signatures, such as the characteristic 10$\mu$
silicate line, have not been generally observed from these shells, it is
believed that the dust is largely carbonaceous.  Further analysis (Ney and
Hatfield 1978; Gehrz, {\it et al.} 1980ab; Bode and Evans 1982; Mitchell,
{\it et al.} 1983) led to typical dust grain diameters of a few tenths
of a micron or, in some models, as large as a few microns, and to visible
optical depths $\sim 3$.  However, extensive modeling of
grain nucleation and growth, for example by Yamamoto and Hasegawa (1977) and
Draine and Salpeter (1977), has not been able to reproduce the grain size
distributions that the emission models appear to require.

Donn and Nuth (1985) have given several reasons for why classical nucleation
theory is inapplicable in many astrophysical environments.  Classical
nucleation
theory uses macroscopic parameters like surface tension to calculate the free
energy of tiny clusters, and assumes the gas temperature and the cluster
vibrational temperature are equal.  In fact, cluster temperatures are usually
controlled by the radiation field, and do not equal the gas temperature.
The theory assumes in its derivation that very small clusters are
present in equilibrium concentrations.  This is incorrect around novae, because
the densities are so small and conditions change so quickly.  Carbon monomers
may have a number density of $10^7$ cm$^{-3}$ and a thermal velocity of $4
\times 10^5$ cm sec$^{-1}$, a collision cross-section of $4 \times 10^{-16}$
cm$^2$, and a mean time between collisions of $\approx 600$ seconds.
With a sticking coefficient of $\approx 10^{-6}$ (Draine 1979), the average
carbon atom would become part of a carbon dimer in $6 \times 10^8$ seconds.
The dust formation process takes place within $3 \times 10^6$ seconds, so
dimers will not be in equilibrium with monomers.  Classical nucleation
theory is therefore inapplicable to nova ejecta.

Because nucleation theory is inapplicable and fails to account for the
observations, we have performed kinetic calculations of the growth of small
condensation nuclei.  Calculations using theoretical sticking probabilities
on small clusters led to grain sizes which were far too small to explain the
infrared and visible data.  Too many clusters grow, and the carbon vapor is
depleted before they reach the empirically inferred size.  We then realized
that the presence of a significant ultraviolet flux would efficiently
photodissociate small grains.  The absorption of even a single
photon raises the temperature of a small cluster sufficiently to induce
evaporation of a carbon atom, a process analogous to photodissociation of a
molecule.  This offered hope of avoiding the excess of
condensation nuclei.  Our calculations showed that this destruction
mechanism is, indeed, too efficient, and that it effectively prevents
the formation of any grains at all.  It is possible that processes we have
not included, such as the effects of hydrogenation of the grain surfaces,
are crucial, or that grains form heterogeneously on pre-existing nuclei.
\bigskip
\centerline{II. The Nucleation Model}
\medskip
In our study of nucleation around novae we integrated the following
set of 60 equations using the IMSL routine DGEAR:  A volume $V$ contains
$N_g$ clusters of size $g$, each cluster containing $f(g)$ carbon atoms.
We chose $f(g) = g$ for $1 \le g \le 20$, $f(g) = 20 \times 2^{(g-20)/5}$
for $21 \le g \le 24 $, and $f(g) = 20 \times 2^{g-24}$ for $g \ge 25$.
We describe clusters of carbon atoms from monomers up to grains with $20
\times 2^{36} \approx 1.37 \times 10^{12}$ atoms, or grains with a radius
$\approx 1 \mu$.  The evolution equations for $3 \le g \le 60$ are
$$\eqalign{{dN_g \over dt} &= {v_{1,g-1} \alpha_{g-1} A_{g-1} N_{g-1} N_1 \over
(f(g)-f(g-1)) V(t)} - {v_{1,g} \alpha_g A_g N_g N_1 \over (f(g+1)-f(g))V(t)}
\cr &\qquad + {v_{2,g-1} A_{g-1} N_{g-1} N_2 \over (f(g)-f(g-1)) V(t)} -
{v_{2,g} A_g N_g N_2 \over (f(g)-f(g-1)) V(t)} \cr &\qquad - {\beta_g N_g \over
f(g)-f(g-1)} + {\beta_{g+1} N_{g+1} \over f(g+1) -f(g)}.\cr} \eqno(1)$$
For $g = 2$
$$\eqalign{{dN_2 \over dt} &= -\sum_{g=3}^{60} {v_{2,g} A_g N_g N_2 \over
(f(g)-f(g-1)) V(t)} - {2 v_{2,2} A_2 N_2^2 \over V(t)} \cr &\qquad - {\alpha_2
v_{1,2} A_2 N_2 N_1 \over V(t)} + \beta_3 N_3 - \beta_2 N_2 +
{\alpha_1 v_{1,1} A_1 N_1^2 \over V(t)}.\cr} \eqno(2)$$
For $g = 1$
$${dN_1 \over dt} = - \sum_{g=2}^{60} f(g) {dN_g \over dt}. \eqno(3)$$
Here $A_g$ and $\alpha_g$ are the cross-section and sticking coefficient
respectively for an atom colliding with a $g$-cluster.  For collisions
of dimers with larger clusters
the sticking coefficient is assumed to be 1; the dimer is assumed to
split and the excess energy to be carried off by one atom which does not
stick to the cluster, while the cross-section is taken to be the same as
that for atoms.  The evaporation coefficient for g-clusters is $\beta_g$
and includes both ordinary thermal evaporation and the evaporation
which takes place immediately after a photon is absorbed.
$V(t)$ is the (arbitrary) volume of the volume element we examine and is
initially equal to 1 cm$^3$ when we start the calculation.
$v_{1,g}$ is the mean relative velocity of a $g$-cluster and a monomer, while
$v_{2,g}$ is the mean relative velocity of a $g$-cluster and a dimer.

The collision cross-section is taken as $f(g)^{2/3} \pi a_0^2$, where
$a_0$ is the radius of a carbon atom.  The sticking coefficient
was calculated as suggested by Williams (1972), Donn {\it et al.} (1981),
and Freed {\it et al.} (1982): When two atoms or an atom and a cluster
collide they form a new, larger cluster with more than enough energy to
dissociate.  The excess energy must be disposed of somehow or the new
transient cluster will break up.  In terrestrial environments the density
of atoms or molecules is high enough that a third body may collide
with the excited cluster and carry off the excess energy.  Around novae
such three body reactions are very unlikely.  Collisions with hydrogen may
occur on a time scale of $(n v A)^{-1}$ where $n \approx 10^9$, $v \approx
10^6$ cm sec$^{-1}$, and $A \approx 10^{-15}$ cm$^2$; the time scale is
$\sim 1$ second.  An excited dimer will only last about $10^{-13}$ second,
so that only one collision in $10^{13}$ will be stabilized by a third body.

In such low density environments stabilization is more likely to occur
by the emission of a photon which can carry off the excess energy.  Such
reactions are called radiative association reactions.
The sticking coefficient, called the probability of radiative stabilization by
Donn, {\it et al.} (1981), is
$$\alpha = {\tau_r^{-1} \over \tau_r^{-1} + \tau_d^{-1}}, \eqno(4)$$
where $\tau_r$ is the radiative lifetime of the excited cluster
and $\tau_d$ is its dissociative lifetime.
We need to determine these two quantities.

The radiative lifetime $\tau_r$ is $\sim 10^{-7}$ second if emission
takes place by an allowed electronic transition, and is $\sim 10^{-2}$
second if cooling takes place by an allowed vibrational transition.  Clearly
sticking is more likely if electronic transitions occur.  For the formation
of dimers $\tau_d$ is equal to a vibrational period, or $\sim 10^{-13}$
second.  Then for C + C $\to$ C$_2$,\break the sticking coefficient is
either $\sim 10^{-6}$ or $\sim 10^{-11}$, depending on which sort of
emission process takes place.  The empirical sticking coefficient of
$\sim 10^{-6}$ (Draine 1979) implies that an electronic transition occurs
(also, vibrational transitions are forbidden for C$_2$ and their actual rate
is much slower than $10^2$ sec$^{-1}$).  We generalize from dimers by using
the radiative lifetime of excited electronic states in calculating all
sticking coefficients.  If this is incorrect it will overestimate the
sticking coefficient.

The other time scale is $\tau_d$, the average lifetime of the excited cluster
before it breaks up.  We estimate this using RRK theory (Johnston 1966,
Duley and Williams 1984), which assumes that the cluster is a collection
of harmonic oscillators of some frequency $\nu$,
which for carbon clusters we take to be $4.5 \times 10^{13}$ sec$^{-1}$
(Freed {\it et al.} 1982).  One particular bond (or vibrational normal
mode) is assumed to be more likely than the others to break up, and if this
mode
accumulates enough energy dissociation occurs.  The internal energy of the
cluster is distributed among the various oscillators in packets of magnitude
$h\nu$.  The dissociation energy $E_d = Mh\nu$.  The total energy
of the cluster is $E = Jh\nu$.  The number of oscillators $S$ is equal to the
number of vibrational modes of the cluster: $S = 3N - 6$ for $N \ge 3$
(we take these molecules to be nonlinear, although in fact C$_3$ is linear)
and $S = 1$ for dimers.  The
number of ways to distribute $J$ identical packets or quanta of vibrational
energy among $S$ modes is $(J+S-1)! \over J! (S-1)!$.  For dissociation to
occur at least $M$ quanta must be in one particular vibrational mode.
Then the number of ways to distribute the remaining $J-M$ quanta over
the $S$ oscillators is $(J-M+S-1)! \over (J-M)! (S-1)!$.  The fraction of
of the total number of distributions which will lead to dissociation is
$(J-M+S-1)! J! \over (J+S-1)! (J-M)!$.  If rearrangements of quanta occur
at a rate $\nu$, the dissociation rate is
$$\tau_d^{-1} = \nu {(J-M+S-1)! J! \over (J+S-1)! (J-M)!}. \eqno(5)$$
If there are $q$ modes which lead to dissociation, we multiply this expression
by $q$.  For dimers, $S = 1$ and $\tau_d^{-1} = \nu$.

To calculate the sticking coefficient we write the
mean energy of a cluster which has just been hit by a gas-phase monomer
with the mean kinetic energy $1.5 k_B T_{gas}$:
$$E = S k_B T_{grain} + 1.5 k_B T_{gas} + E_d, \eqno(6)$$
where $T_{grain}$ is the grain temperature, $T_{gas}$ is the gas
temperature, $k_B$ is Boltzmann's constant, and $E_d$ is the dissociation
energy which we take to be 6 eV.
We then have $J = E/h\nu$ and $M = E_d/h\nu$, and can calculate
$\tau_d^{-1}$.  The sticking coefficient $\alpha$ is shown in Table 1.

The sticking coefficient falls short of 1, even for very large clusters,
because then $J \approx S (k_B T_{grain}/h\nu) \gg M $ and
$$\tau_d^{-1} \to \nu \left({J \over J+S-1}\right)^M \to \nu \left({k_B
T_{grain} \over k_B T_{grain} + h \nu}\right)^M.  \eqno(7)$$
This is much less than $\nu$, but it is not zero.

The equation for relative velocity is taken from Draine and Salpeter (1979),
who give an expression for the drag force $F_{drag}(v)$ on a neutral grain
moving through a gas.  We balance this
velocity dependent drag force against the force of radiation pressure:
$$F_{rad} = {L \over 4 \pi r^2 c} Q \pi a^2, \eqno(8)$$
where $a$ is the grain radius, $L$ the nova's luminosity, and we take the
effective momentum transfer coefficient $Q = \min(10^5a,1)$.  Equating
$F_{drag}(v_{gr})=F_{rad}$ gives $v_{gr}$, the velocity of a grain through the
gas.  We can then calculate the flux of carbon atoms hitting a grain using
the equations derived by Aannestad and given by Shull (1978).  The result is
$$F = \left({\pi \over 2}\right)^{1/2} a^2 n_C {v_{th} \over v_{gr}} \left[
(2\pi)^{1/2} v_{th} \left(1 + {v_{gr}^2 \over v_{th}^2}\right)\,{\rm erf}\,
\left({v_{gr} \over 2^{1/2} v_{th}}\right) + 2 v_{gr} \exp \left(-{v_{gr}^2
\over 2 v_{th}^2}\right) \right], \eqno(9)$$
where $v_{th} \equiv (k_B T_{gas}/m_C)^{1/2}$ is the atomic thermal velocity
and $n_C$ is the density of carbon atoms.  When $v_{gr} \gg v_{th}$ this
reduces to
$$F = \pi a^2 n_C v_{gr}, \eqno(10)$$
as it must.

For our system of grain-growth equations (1)--(3), we define the relative
velocity $v$ so that $n_C v \pi a^2$ gives the flux.  Then
$$v = {F \over \pi a^2 n_C}. \eqno(11)$$
We calculate this velocity for collisions between monomers, dimers and
larger clusters.  For the relative velocities of two monomers or
monomers and dimers, we used $v = \left({8 / \pi}\right)^{1/2}v_{th}$,
which is not strictly correct, but adequate given the uncertainty in the
reaction cross-section.  For dimer-dimer collisions we similarly used
$v = \left(4 / \pi\right)^{1/2}v_{th}$.

We split the evaporation rate into two terms, one giving the evaporation
rate at the cluster temperature $T_m$ corresponding to the mean heat flux,
and the other giving the evaporation rate of the cluster as a result of
absorption of single ultraviolet photons (photodissociation).  This is a
somewhat artificial distinction, because the temperature of the cluster is
determined by a balance of photon absorption and infrared emission.  For very
small clusters the temperature fluctuations are much larger than the average
temperature, while for clusters composed of many atoms the absorption of a
single ultraviolet photon has little effect on the temperature.  The method
outlined in the Appendix treats dissociation in a unified way.  Here we
describe a simpler approach.

$T_m$ is determined by the equation
$$T_m = \left({L \over 16 \pi \sigma r^2}{Q_{UV} \over Q_{IR}}\right)^{1/4},
\eqno(12)$$
where the ratio of emissivities $Q_{UV}/Q_{IR} = (T_*/T_m)^{1.65}$ (Clayton
and Wickramasinghe 1976), $\sigma$ is the Stefan-Boltzmann constant, and
$T_*$ is the color temperature of the nova.  We adopt $T_* =
8000^{\,\circ}$K.  Then the evaporation rate of a spherical cluster is
$$\beta = 4 \pi a^2 \alpha \left({P_s \over (2 \pi m_C k_B T_m)^{1/2}}\right)
\exp\left({2 \gamma \Omega \over a k_B T_m}\right), \eqno(13)$$
where the exponential term comes from the Gibbs correction to the vapor
pressure,\break $P_s = 1.66 \times 10^{14} \exp(-88880^{\,\circ}{\rm K}/T_m)$
(Lefevre 1979) is the saturation vapor pressure over a flat surface,
$\alpha$ is the sticking coefficient,
$\Omega$ is the volume of an atom in the condensed phase, and $\gamma =
1000$ erg cm$^{-2}$ is the surface energy per unit area.  For large grains
the mean evaporation rate (13) exceeds that resulting from fluctuations.

For small clusters fluctuations dominate.  A cluster with 4 atoms has 6
degrees of freedom.  A 6 eV photon will give an average of 1 eV of energy to
each degree of freedom, increasing its temperature by 11000$^{\,\circ}$K.
We then must calculate the probability of dissociation occurring while
this high temperature lasts---the photodissociation rate.
\bigskip
\centerline{III. Photodissociation}
\medskip
In order to calculate the photodissociation rate we must calculate the
probability that a cluster reradiates its absorbed energy before it
dissociates.  If the average radiating time is $\tau_r$ and the average
dissociating time is $\tau_d$, then the chance of its dissociating before
it loses energy by radiation is $\tau_d^{-1} / (\tau_d^{-1} +
\tau_r^{-1})$.  The dissociation rate from RRK theory is given by (5).
This rate has its smallest value if $E = E_d$ ($J = M$).  Then
$$\tau_d^{-1} = \nu {(S-1)! M! \over (M+S-1)!}. \eqno(14)$$
The larger the cluster the more degrees of freedom it has, and the less likely
it is to dissociate.  The probability of breakup is
$$P = {\tau_d^{-1} \over \tau_r^{-1} + \tau_d^{-1}}. \eqno(15)$$

Omont (1986) describes what happens when a polycyclic aromatic hydrocarbon
mole\-cule, which resembles a small carbon cluster, absorbs an ultraviolet
photon.  If the energy is great enough for photoionization, ionization can
occur with a probability $>0.5$.  Photodissociation can also occur immediately,
if the excited electronic state is vibrationally unbound.  If neither of these
events occurs, the energy generally is thermalized as vibrational energy
(reradiation by an electronic transition is possible, and is accounted for
by use of an empirical absorption cross-section).  The cluster then cools
by vibrational infrared radiation or by evaporation.

We show in Table 2 dissociation times $\tau_d$ of pure carbon clusters,
assuming that their internal energy exactly equals the dissociation energy,
taking $\nu = 4.5 \times 10^{13}$ sec$^{-1}$, $M=30$, and $S = 3N-6$, and
using (14).  Clusters of $N < 8$ will readily photodissociate.  Unless the
ultraviolet spectrum is very hard, only one carbon atom will be lost for
each absorbed photon with $h \nu_{UV} > E_d$.

It is important to know at what wavelengths the clusters in question can
absorb.  A well-known paper by Platt (1956)
estimated the longest wavelength which could be absorbed by a molecule with
unsaturated energy bands.  Platt treats the molecule as a rectangular box
with dimensions equal to the dimensions of the molecule.  Each atom is assumed
to contribute one electron and the lowest energy levels are filled up.  The
amount of energy it takes to raise an electron from the highest filled energy
level to the next one above it gives, in this model, the lowest energy photon
which the molecule can absorb.  For linear molecules the wavelength of the
photon is about 400 times the length of the molecule.  Platt found that
experimental data for many different organic molecules agreed with this
prediction to within a factor of two.  Thus, molecules about 10\AA\ long
are just barely able to absorb in the visible region.  Certainly even very
small molecules will be able to absorb ultraviolet light.  We will ignore
absorption by visible photons, which have energies between 2 and 4 eV.  Two or
more successive absorptions of these photons would provide enough
energy to break loose a carbon atom, but unless the absorption rate is very
high the molecule will reradiate much of its energy in the interval between
absorptions; hence we ignore this route to photodissociation.

Having decided only to include ultraviolet photons in our simulation, we must
assign the small clusters an absorption cross-section.  From Lee (1984), it
appears that for many small molecules the absorption cross-section is roughly
$10^{-18}$ cm$^2$.  Omont (1986) gives a value of $\approx 5 \times 10^{-19}$
cm$^2$ per carbon atom for the absorption cross-section of polycyclic aromatic
hydrocarbons at visible wavelengths; for the ultraviolet wavelengths the
cross-sections are an order of magnitude larger.  We adopted a value of $1.5
\times 10^{-18}$ cm$^2$ atom$^{-1}$.
\bigskip
\centerline{IV. Results}
\medskip
We first describe calculations without photodissociation.  The evaporation
rate was given by (13), using the mean temperature (12).  In one calculation
we assumed that $10^{27}$ gm of carbon atoms were present and spread evenly
throughout a shell of volume $4 \pi r^3 / 10$;  {\it i.e.}, at a given
instant there was a shell of radius $r = vt$ and of thickness $\approx
r/10$.  We began the calculation at a time after the nova outburst
$t = 10^6$ sec, and the expulsion velocity $v = 750$ km sec$^{-1}$.
Nucleation begins around $t =  2.5 \times 10^6$ sec.
By $t =  3.4 \times 10^6$ sec less than 1\% of the carbon remains
monomeric.  The distribution of grain sizes is shown by the solid line in
Figure 1.  Most of the mass is in grains with $a \approx 100$ \AA,\break in
disagreement with the infrared data.  The visible optical depth (calculated
using the opacity of Clayton and Wickramasinghe 1976) is 90, much greater
than observed.  If the shell were 30 times thinner (but of the same density)
and
the carbon mass were 30 times less, the optical depth would be 3, as observed
for Nova Vul 1976, Nova Ser 1978, and Nov Ser 1970.  Such a thin shell,
equal to $r/300 \sim 10^{12}$ cm in thickness, is unlikely to form and would
not remain thin for a period greater than $\Delta r / 2 v_s$
(Ennis, {\it et al.} 1977), where $\Delta r$ is its thickness and $v_s$
is its sound speed.

In another calculation the carbon mass was arbitrarily increased to $10^{28}$
gm in order to see if larger grains would result.  Most of the
carbon formed dust by $t = 2.6 \times 10^6$ sec.  The distribution of grain
sizes is shown by the dashed line in Figure 1.  Most of the carbon is in
grains with radii $a \approx 0.15\mu$, roughly 10 times greater than
before.  This is perhaps large enough to agree with the infrared data.
However, the visible optical depth was $\approx 700$, which is grossly
excessive.  To keep the same density of carbon, and
therefore the same grain size, but to have the right visible optical depth, we
must reduce the mass and shell thickness by a factor of roughly 300, so that
$\Delta r = r/3000 = 10^{11}$ cm.  This is
far too small.  We have the same problem that classical nucleation theory
has---to produce large grains ($a > 0.1\mu$) we require a very high density
of carbon.  The resulting visible optical depth is too great.

It might be that high energy photons keep the number of nucleation sites well
under the number predicted by our model.  If so, the carbon might accrete
onto relatively few nucleation sites to form larger grains.  This is
possible, if the nucleation sites are sufficiently rare, because once a
nucleus overcomes the barrier at $N \le 7$ it grows rapidly.  It was with
hopes that this might prove to be the case that we included
photodissociation in the calculation.  We assumed a blackbody nova spectrum
with $L = 1.25 \times 10^{38}$ erg sec$^{-1}$ and $T_* = 8000^{\,\circ}$K.
All photons with 6.0 eV $< h \nu_{UV} <$ 13.6 eV were assumed to
contribute to the photodissociation rate, although in using RRK theory
(5) to describe photodissociation we took each to have an energy $h
\nu_{UV} = 6.5$ eV, only slightly above the assumed photodissociation
threshold $E_d = 6$ eV.

In separate calculations
we took the radiative decay time of the excited state as $10^{-7}$
second, characteristic of an excited electronic state, and $10^{-2}$
second, characteristic of excited vibrational states.  The longer radiative
time is appropriate (Omont 1986) because the RRK theory applies to
vibrationally excited molecules.  The choice of radiative times for the
molecule shifts the dividing line between
small clusters which are almost certain to dissociate, and large clusters
which are are almost certain not to dissociate.

We assumed that $10^{28}$ gm of carbon (corresponding to $10^{-4} M_\odot$
of ejecta, of which 5\% by mass is carbon) were spread throughout a shell of
volume $4 \pi r^3 / 10$, which is expanding at a speed of 750 km/sec.
The calculation was begun at the time $t = 10^6$ sec, and resembled our
previous calculation except for the inclusion of photodissociation.
The results are dramatically different in that no dust forms for either
value of the radiative decay time.  Virtually no dimers form; with $\tau_r =
10^{-2}$ second, at the calculation's end the ratio of dimers to monomers
is $6 \times 10^{-10}$.  Essentially all the carbon remains as monatomic
vapor.

The bottleneck to grain formation occurs at very small
molecules: C$_2$, C$_3$, C$_4$, {\it etc.}, for which absorption of a 6.5 eV
photon virtually guarantees dissociation, since $\tau_d$ is less
than $10^{-7}$ second for these tiny clusters.  It does not matter whether
$\tau_r$ is $10^{-7}$ second or $10^{-2}$ second, since the clusters for
which $\tau_d$ is comparable even to $10^{-7}$ second form in only
extremely tiny numbers, although the larger (and physically expected)
$\tau_r$ worsens the problem.

We tried again with $\tau_r = 10^{-2}$ second for the clusters and $10^{28}$
gm of carbon spread throughout a volume of $4 \pi r^3 / 100$.  The
radiation rate is realistic, but such a thin shell, with thickness $\approx
r/100$, is not.  It does give a very high carbon density.  Nonetheless, no
nucleation occurs.  The ratio of dimers to monomers is increased tenfold, as
one expects.  Still the number of dimers is only $6 \times 10^{-9}$ of the
number of monomers and the number of trimers is $10^7$ times rarer still.

We attempted one more run, with a carbon mass of $10^{28}$ gm, a shell
volume of $4 \pi r^3 / 10$, an expulsion velocity of 750 km sec$^{-1}$, and
$\tau_r = 10^{-2}$ second.  In this last case, for our
photodissociation rate we used the same equation (13) that we used for the
evaporation rate, with $T$, the ``temperature'' of the molecule set equal to
$E_p/S$.  Here $E_p$ is the energy of the photon, set equal to 6.5 eV,
and S is the number of degrees of freedom, and equals 3N-6 (or 1, for dimers).
The formula has the advantage of being consistent with the equation we use to
calculate the evaporation rate of the clusters at a constant temperature.  It
does not really matter, for the photodissociation rate for the small clusters
is identical to the rate one finds with RRK theory.  It merely equals the
photon absorption rate.  The results are then of course the same:  no
nucleation and no dust.

No dust forms around novae in our nucleation calculations because
ultraviolet photons disrupt small clusters of carbon atoms before they can
grow.
The exact method we use to determine the rate of dissociation of an excited
cluster is not important, because the photodissociation rate of small clusters
equals their photon absorption rate.
\bigskip
\centerline{V. Discussion}
\medskip
A simple model can be constructed to show how the photodissociation rate
affects nucleation.  Let the equation for the number density of $i$-mers
$n_i$ be written as
$${dn_i \over dt} = R_{i-1} n_{i-1} n_1 -k_i n_i - R_i n_i n_1 + k_{i+1}
n_{i+1}, \eqno(16)$$
where $R_i = v \sigma_i \alpha_i$.  For simplicity we set $v \sigma_i =
10^{-10}$ cm$^3$ sec$^{-1}$ for all small clusters.  For the sticking
coefficient we take $\alpha_i = \min(10^{-7+i},1)$.  This approximation is
slightly more optimistic than that used by Donn, {\it et al.} (1981) and
than the numbers given by the RRK theory.  For $k_i$ we assume a
photodissociation cross-section of $10^{-18}$--$10^{-17}$ cm$^2$.  The
number of photons with $E_p > 6$ eV emitted by a nova with $L = 1.25 \times
10^{38}$ erg sec$^{-1}$ depends on its effective temperature and is in the
range $10^{47}$--$10^{49}$ sec$^{-1}$.  At a distance of $3 \times 10^{14}$
cm the flux will be $10^{17}$--$10^{19}$ photons cm$^{-2}$.  Then the $k_i$
are in the range $10^{-1}$--$10^2$ sec$^{-1}$.

For $n_1$ we have a range to pick from.  Let the carbon mass be $10^{27}$--$
10^{28}$gm.  Then, given a shell of volume $4 \pi r^3 / 10$, the carbon
density at $r = 3 \times 10^{14}$ cm is $1.6 \times 10^6$--$1.6 \times 10^7$
cm$^{-3}$.  A $10^{28}$ gm shell of thickness $r/100$ with clusters forming at
$r = 10^{14}$ cm gives $n_1 = 5 \times 10^9$ cm$^{-3}$.  We assume
(optimistically) that C$_6$ is immune to photodissociation (hydrogenated
C$_6$ may be the stable benzene molecule); once a
cluster reaches C$_6$ it forms a nucleation site.  In reality it
must overcome more hurdles and grow still larger to be completely safe from
destruction from photodissociation.  Then the nucleation
rate must be less than $R_5 n_5 n_1$, the formation rate of C$_6$.

To obtain useful analytic results from these equations we must assume that a
quasi\-equilibrium state is reached so that ${dn_i \over dt} = 0$.  This is not
a reasonable assumption to make when there is no photodissociation,
as Yamamoto and Nishida (1977) demonstrated.  It is reasonable in our
case: In equilibrium $n_{i+1} \ll n_i \ll n_{i-1}$.
Then, neglecting the last two terms on the right hand side of (16),
$$n_i \approx {R_{i-1} n_{i-1} n_1 \over k_i}. \eqno(17)$$
Later, when we substitute numbers for $n_1$, $R_{i-1}$, and $k_i$, we will
see that our assumption that $n_i / n_{i-1} \ll 1$ is justified.  The
equilibrium value is approached on a time scale $O(k_i^{-1})$.  Now $k_i$ is
between 0.1 and 100 sec$^{-1}$, so that it takes no more than $\sim 10$
seconds to reach equilibrium.  The expansion time scale (the time in which
a significant change in the volume of the ejecta takes place) is much
greater, justifying the use of the quasiequilibrium distribution.

We can rewrite the equilibrium distribution as
$$n_i \approx {R_{i-1} R_{i-2} \ldots R_1 n_1^i \over k_i k_{i-1} \ldots k_2}.
\eqno(18)$$
Then the nucleation rate $J < R_5 n_5 n_1$, or
$$J < {R_5 R_4 R_3 R_2 R_1 n_1^6 \over k_5 k_4 k_3 k_2}. \eqno(19)$$
Our previous estimate for the $\alpha_i$ gives $R_i = 10^{-17+i}$ cm$^3$
sec$^{-1}$ ($i \le 7$), and we set $k_i = k$, so that all of the small
clusters have the same photodissociation rate (roughly justified because the
photodissociation probability is nearly unity).  Then
$$n_i \approx {10^{-17(i-1)} 10^{(i^2-i)/2} n_1^i \over k^{i-1}}
{\rm cm}^{3(i-1)}{\rm sec}^{-(i-1)}, \eqno(20)$$
or
$$n_5 \approx 10^{-58} {\rm cm}^{12} {\rm sec}^{-4} {n_1^5 \over k^4}
\eqno(21)$$
$$J < 10^{-70} {\rm cm}^{15} {\rm sec}^{-5} {n_1^6 \over k^4}. \eqno(22)$$
In $\Delta t = 10^6$ sec, there will be $J \Delta t$ nucleation
sites produced per unit volume, or about $\xi \equiv J \Delta t / n_1$
per carbon atom.  This is the maximum
possible number of nucleation sites that could be produced
during the nova eruption under these assumptions.  Then $\xi < 10^{-64}
{\rm cm}^{15} {\rm sec}^{-4} n_1^5/k^4$.

If at least 10\% of the carbon condenses, we require $\xi \sim 10^{-10}$
to obtain 0.1$\mu$ grains, or $\xi \sim 10^{-13}$ to form 1$\mu$ grains.
For plausible values such as $n_1 = 10^8$ cm$^{-3}$ and $k = 1$ sec$^{-1}$
we find $\xi < 10^{-24}$ and no dust forms, just as we found from numerical
integration of the equations (1)--(3).  If $n = 10^9$ cm$^{-3}$ and $k = 0.1$
sec$^{-1}$, which are very optimistic, $\xi < 10^{-15}$ and very little dust
will form.  Only for extreme values $n_1 = 10^{10}$ cm$^{-3}$ and $k = 0.1$
sec$^{-1}$ can $\xi$ be as large as $10^{-10}$.
This corresponds to $10^{27}$--$10^{28}$ gm of carbon compressed into a volume
of $5 \times 10^{39}$--$5 \times 10^{40}$ cm$^3$, {\it i.e.}, a shell of
radius  $10^{14}$ cm and thickness $10^{11}$--$10^{12}$ cm.  Alternatively,
one could assume a clump of gas $\approx 10^{13}$ cm in radius which forms
the dust.  This would require very asymmetric mass ejection.  Mitchell {\it
et al.} (1985) rejected this possibility.  Even if we have extremely high
densities, the photodissociation rate must be $< 10^{-1}$ sec$^{-1}$ for
dust to form.  For a black-body source of $L = 1.25 \times 10^{38}$ erg
sec$^{-1}$ at $r = 3 \times 10^{14}$ cm and a photodissociation
cross-section of $10^{-18}$ cm$^2$ this would require $T_* <
7000^{\,\circ}$K.
\bigskip
\centerline{VI. Conclusions}
\medskip
In our nucleation calculation we showed that it is difficult to understand
how nucleation can take place at all.  The first obstacle is photoionization;
if the radiating temperature of the nova is greater than 12000$^{\,\circ}$K it
is likely that the carbon is entirely ionized and no nucleation can take place.
Assuming that in the early stages of the eruption the nova is cooler than this,
there are still a great many photons in the 6-11 eV range, below the carbon
ionization threshold but still capable of
photodissociating small molecules and clusters.  They prevent nucleation.

There may be ways around the photodissociation problem.  Some of the effects
of hydrocarbon chemistry on dust nucleation around novae were discussed by
Rawlings and Williams (1989).  Omont (1986)
pointed out that excited polycyclic aromatic hydrocarbons are more likely to
lose a hydrogen atom than a carbon atom, and this
will also be true for grain nuclei with any hydrogen attached.  Hydrogen
is more loosely bound and it is thus more likely to accumulate the necessary
amount of energy to break away.  The remaining energy will usually be too low
for any remaining atoms to escape.  Using RRK theory (5) we see that a
hydrogen atom is $\zeta$ times more likely to break off, where
$$\zeta = {(J-H+S-1)! (J-C)! \over (J-H)! (J-C+S-1)!}, \eqno(23)$$
$H h \nu$ is the binding energy of a hydrogen atom, and
$C h \nu$ is the binding energy of a carbon atom.  If, for instance,
$C = 30$, $H = 20$, $J = 30$, and $S = 6$, a given hydrogen atom is 3000
times more likely to be lost than a given carbon atom.  This ratio drops
steeply with increasing $J$ (for $J = 35$ it is only 60) and decreasing $S$,
but a large effect is possible.

We neglected hydrogen in our calculations of grain growth, but its number
density in the nova debris is typically 100--1000 times that of carbon,
and 300--3000 hydrogen atoms collide with a cluster for every carbon atom
which does so.  Any
carbon cluster that does form will likely be hydrogenated on the outside.
Collision with a carbon atom will exothermically replace a hydrogen by the
carbon; the presence of hydrogen available to carry off the carbon's
attachment energy will increase the sticking probability of carbon as well
as reducing the carbon photodissociation rate.  Frequent collisions with
hydrogen atoms will replenish the hydrogen layer, and will increase the rate
of carbon accretion as well as protecting the nascent cluster from
photodissociation.  The rate of growth of a carbon cluster will then be
limited by the rate at which it can capture the more abundant hydrogen.
Further calculations are needed to quantify this effect.

On the other hand, if nucleation is impossible around an erupting nova, we
must explain why dust seems to form when the ejected material has had time to
reach the condensation radius.
It may be that pre-existing nucleation sites lie around the nova before the
eruption, too small to obstruct much light unless they grow (Malakpur 1977,
Bode and Evans 1980, Stickland, {\it et al.} 1981, Albinson and Evans 1987;
but see objections by Bode 1982).  When the eruption begins, the nuclei
inside the condensation radius evaporate, while those outside may survive
the initial encounter with ejecta and begin to grow.  Alternatively, nuclei
may form in the dense gas of an accretion (or exterior excretion) disc, where
they are shielded from direct irradiation by the nova.  They may then be swept
into the outward gas flow.  Hypotheses of this type depend sensitively on
details of mixing in complex fluid flows.

We thank S. H. Margolis for useful discussions and NASA NAGW-2918 for
partial support.
\vfil
\eject
\centerline{Appendix: A Stochastic Approach to Cluster Evaporation}
\medskip
In the body of this paper we calculated the photodissociation rate of a
small cluster assuming it absorbs only one photon.  When the photon flux is
high this assumption is not justified, for the cluster has a significant
chance of absorbing a second photon before it loses all the energy of the
first.  This effect is unimportant for our grain growth calculations, for
under such high photon fluxes photodissociation destroys all small particles,
and no larger ones form.  It may, however, be of interest for the evolution
of larger grains in the presence of large photon fluxes and for the growth of
pre-existing nuclei.  It also provides a mechanism by which photons whose
energy is individually less than the photodissociation threshold may cause
photodissociation.  In this Appendix all photons are assumed to have the
same energy of 4 eV, a fair approximation for grains heated by an
$8000^{\,\circ}$K black body nova spectrum, but the dissociation energy is
still taken to be 6 eV.

The heating of a cluster by visible or ultraviolet radiation can be treated
as a shot noise process, with the absorption of photons being a Poisson process
and their cooling by infrared emission determining the response function.  The
cooling rate is directly proportional to the energy content of the cluster, so
that its response is linear.  The response function is $E_p \exp(-bt)$,
where $E_p$ is the energy of the absorbed photon and $b = 100$ sec$^{-1}$,
appropriate to infrared emission by molecular vibrations.  We need to
determine the probability distribution of the internal energy of the cluster,
given an average photon absorption rate $\lambda$ and a cooling time $1/b$.

The solution to this statistical problem
was given by Papoulis (1966), under the approximation that the cooling
after a photon absorption follows an exponential law for a finite time $T$
and that the energy then abruptly drops to zero.  This approximation becomes
exact in the limit $T \to \infty$.  The probability density $f(E)$ is
$$f(E) = \exp{(-\lambda T)} \sum^\infty_{k=0} g_k(E) {(\lambda T)^k \over k!},
\eqno(A.1)$$
where $g_k(E)$ is the probability density of the cluster's energy,
given that it has absorbed $k$ photons in the time interval $T$.

The function $g_0(E) = \delta(E)$
and the other $g_k(E)$ can be derived in the following way:
If a photon of energy $E_p$ was absorbed in the preceding interval $T$ its
remaining contribution to the cluster will be an energy between $E_p$ and
$E_p \exp(-bT)$; if it was absorbed a time $t$ ago its remaining contribution
is $E_p \exp(-bt)$.  For any $0 < t < T$  the probability
$P(E)$ that such a photon contributes less than $E = E_p \exp(-bt)$ is
$(T - t) / T = 1 + \ln(E/E_p)/(bT)$.  The probability density $g_1(E)$ is
$$g_1(E) = {dP \over dE} = {1 \over bTE}. \eqno(A.2)$$
The probability density $g_2(E)$ is the convolution of $g_1(E)$
with itself and, in general,
$$g_k(E) = g_{k-1}(E) \ast g_1(E).  \eqno(A.3)$$
These $g_k$ can be calculated by Fourier transforming $g_1$,
raising it to the $k$-th power, and inverting the transform.

In practice we consider only the first 100 terms in the infinite sum (A.1),
which are sufficient if $(\lambda T)^{100} \ll 100!$, or $\lambda T < 100/e
\approx 37$.
For $T$, we chose a value equal to six half-lives, during which the energy
decays by a factor of 64; {\it i.e.}, $T = \ln(64)/b$.  The introduction of
$T$ is only a calculational device with no physical significance.  Since
the absorbed photons each have an energy
of 4 eV and the emitted infrared photons energies of 0.1--0.2 eV, the
discreteness of the infrared emission, which we ignore, imposes a
physical cutoff at $t = O(T)$.  Our $T$ merely represents the time
it takes for a cluster with 4 eV of internal energy to radiate its last
vibrational quantum.  Because $T$ has no physical significance,
we will present our results in terms of the physically
meaningful term $\lambda \ln2 / b = \lambda T_{1/2}$,
the number of photons absorbed in a cooling half life.

For large fluxes of photons, the probability density $f(E)$ should approach a
Gaussian.  The average cluster energy (in units of 4 eV, the energy of our
photons) is $\lambda /b$.  Its variance is $\lambda/(2b)$ (Papoulis 1966).
Thus we can construct a Gaussian probability
distribution with $\sigma = (\lambda/2b)^{1/2}$:
$$g(E) = {1 \over (2 \pi \sigma^2)^{1/2}}\exp{\left({-(E-\langle E \rangle)^2
\over 2 \sigma^2}\right)}. \eqno(A.4)$$
Given $g(E)$ we can calculate the mean evaporation rate from the evaporation
rate as a function of energy given by the RRK theory of unimolecular
reaction rates (5).  In this appendix we set $S = 3N-5$, which is correct
for the linear dimers and trimers; for nonlinear clusters $S = 3N-6$.
However, the difference is small for the larger clusters; we also
calculated evaporation rates for quadrimers, assuming $S = 3N-6$, and found
evaporation rates increased at most by a factor of 2.5.  For clusters of 20
the difference was less than 30\%.  The uncertainties in the absorption
cross-section, or whether very small clusters can absorb at all at energies
of 4 eV, are of greater importance.

We calculate the mean evaporation rate of clusters containing $N$ atoms and
$S$ modes in a given radiation field using (A.1) and the RRK theory
dissociation rate (5).  It is
$$R = \int^\infty_0 f(E) \tau_d^{-1}(E)\,dE. \eqno(A.5)$$
The radiation intensity enters through the parameter $\lambda$ in (A.1).
This may be used in the kinetic equations for grain growth.

At low values of $\lambda/b$, fluctuations are the dominant effect
controlling the evaporation rate.  If the mean energy is used for the
calculation, clusters whose mean energy is less than the dissociation
energy have an evaporation rate of zero.  The true evaporation rate
can be quite high in small clusters, because of fluctuations which carry
the energy over the dissociation threshold.  At higher fluxes and for
larger clusters (which have a higher absorption rate), the
Gaussian approximation becomes an accurate approximation to the true
evaporation rate.  At still higher flux levels, fluctuations become
insignificant;  the evaporation rates given by the mean energy, the
Gaussian approximation to the energy distribution, and the true energy
distribution all agree.  Figure A.1 displays the regimes in ($\lambda T_{1/2}$,
$N$) space in which the various approximations are valid.  At the highest
$\lambda T_{1/2}$ it is adequate to treat the clusters as having their mean
energy.  At lower $\lambda T_{1/2}$ the Gaussian approximation (A.4) to the
energy distribution is adequate, while for the smallest $\lambda T_{1/2}$ it
is necessary to use the complete stochastic $f(E)$ as in (A.1).

We have calculated evaporation rates for clusters in the size range
2--640 carbon atoms (larger clusters would have high absorption rates and
fluctuations would be unimportant) at distances $r$ in the range $7.5
\times 10^{13}$--$6 \times 10^{14}$ cm from a $L = 10^{38}$ ergs sec$^{-1}$
source of 4 eV monochromatic radiation.  These
parameters describe nova shells at the time when dust is observed to form.
We assume an absorption cross section of $10^{-18}$ cm$^2$ per atom.  Given
these parameters the flux at the dust-forming distance
is about $10^{19}$ photons cm$^{-2}$ sec$^{-1}$.  Dimers
will absorb about 20 photons per second, while an 10-mer will absorb 100
photons per second.  The clusters must begin forming before the dust, and
hence possibly at somewhat smaller $r$ and higher fluxes.
We are therefore interested in a range
of values of $\lambda$ between 10 and $10^3$ sec$^{-1}$, and the time
between photon absorptions is $10^{-3}$--$10^{-1}$ sec.

Some sample results are shown in Table A.1 for $N = 3$ and $N = 20$.
For $\lambda T_{1/2} \le 1$ the average energy of a cluster is less
than 6 eV, the dissociation energy.  The evaporation rate calculated
using the mean energy is zero, while the evaporation rates calculated
for $N = 3$ in the Gaussian approximation and exactly are up to $10^{12}$
sec$^{-1}$, because the fluctuations take some grains above the energy
needed for dissociation.  These extraordinarily high rates are unphysical,
because they violate our tacit assumption that dissociation is rare enough
that the energy distribution of a cluster is determined only by radiative
processes.  They imply actual dissociation rates approaching $\lambda$, at
which this assumption breaks down.  Such dissociation rates exceed the rate
of accretion of carbon atoms and would prevent any cluster
growth.  Our assumptions are valid and our calculational technique applies
when dissociation rates are comparable to association rates, the circumstances
in which quantitative results are needed.

The exact calculation gives a much higher value for the evaporation rate
than the Gaussian approximation at low flux levels; the Gaussian approximation
underestimates the extent of the fluctuations.  As the flux increases the
Gaussian approximation and the exact method become closer; in fact they cross
and the Gaussian approximation becomes slightly larger than the exact approach.
For still higher fluxes the calculation using the mean energy approaches the
other two values.

A similar pattern is followed by clusters of larger size (Table A.2 shows
results for 20-mers).  Here the Gaussian approximation and the exact
calculation come to agree fairly well.  Still higher fluxes are needed to
bring the calculation using the mean energy into agreement with the others.
\vfil
\eject
\centerline{References}
\medskip
Albinson, J. S., and Evans, A. 1987 {\it Ap. Sp. Sci.} {\bf 131}, 443.
\smallskip
Bode, M. F. 1982 {\it Vistas in Astron.} {\bf 26}, 369.
\smallskip
Bode, M. F., and Evans, A. 1980 {\it Astron. Ap.} {\bf 89}, 158.
\smallskip
Bode, M. F., and Evans, A. 1982 {\it MNRAS} {\bf 200}, 175.
\smallskip
Clayton, D. D., and Wickramasinghe, N. C. 1976 {\it Ap. Sp. Sci.} {\bf 42},
463.
\smallskip
Donn, B., Hecht, J., Khanna, R., Nuth, J., and Stranz, D. 1981 {\it Surf. Sci.}
{\bf 106}, 576.
\smallskip
Donn, B., and Nuth, J. 1985 {\it Ap. J.} {\bf 288}, 187.
\smallskip
Draine, B. T. 1979 {\it Ap. Sp. Sci.} {\bf 65}, 313.
\smallskip
Draine, B. T., and Salpeter, E. E. 1977 {\it J. Chem. Phys.} {\bf 67}, 2230.
\smallskip
Draine, B. T., and Salpeter, E. E. 1979 {\it Ap. J.} {\bf 231}, 77.
\smallskip
Duley, W. W., and Williams, D. A. 1984 {\it Interstellar Chemistry} (Academic
Press: London).
\smallskip
Ennis, D., Becklin, E. E., Beckwith, S., Eliot, J., Gatley, I., Matthews,
K., Neugebauer, G., and Willner, S. P. 1977 {\it Ap. J.} {\bf 214}, 478.
\smallskip
Freed, K. F., Oka, T., and Suzuki, H. 1982 {\it Ap. J.} {\bf 263}, 718.
\smallskip
Gehrz, R. D., Grasdalen, G. L., Hackwell, J. A., and Ney, E. P. 1980a {\it
Ap. J.} {\bf 237}, 855.
\smallskip
Gehrz, R. D., Hackwell, J. A., Grasdalen, G. L., Ney, E. P., Neugebauer, G.,
and Sellgren, K. 1980b {\it Ap. J.} {\bf 239}, 570.
\smallskip
Johnston, H. S. 1966 {\it Gas Phase Reaction Rate Theory} (Ronald Press: New
York)
\smallskip
Lee, L. C. 1984 {\it Ap. J.} {\bf 282}, 172.
\smallskip
Lefevre, J. 1979 {\it Astron. Ap.} {\bf 72}, 61.
\smallskip
Malakpur, I. 1977 {\it Ap. Sp. Sci.} {\bf 47}, 3.
\smallskip
Mitchell, R. M., Evans, A., and Bode, M. F. 1983 {\it MNRAS} {\bf 205}, 1141.
\smallskip
Mitchell, R. M., Robinson, G., Hyland, A. R., and Neugebauer, G. 1985 {\it
MNRAS} {\bf 216}, 1057.
\smallskip
Ney, E. P., and Hatfield, B. F. 1978 {\it Ap.J. (Lett.)} {\bf 219}, L111.
\smallskip
Omont, A. 1986 {\it Astron. Ap.} {\bf 164}, 159.
\smallskip
Papoulis, A. 1965 {\it Probability, Random Variables, and Stochastic
Processes} (McGraw-Hill: New York).
\smallskip
Platt, J. R. 1956, {\it Ap. J.} {\bf 122}, 486.
\smallskip
Rawlings, J. M. C., and Williams, D. A. 1989, {\it MNRAS} {\bf 240}, 729.
\smallskip
Shull, J. M. 1978 {\it Ap. J.} {\bf 226}, 858.
\smallskip
Stickland, D. J., Penn, C. J., Seaton, M. J., Snijders, M. A. J., and Storey,
P. J. 1981 {\it MNRAS} {\bf 197}, 107.
\smallskip
Williams, D. A. 1972 {\it Ap. Lett.} {\bf 10}, 17.
\smallskip
Yamamoto, T., and Hasegawa, H. 1977 {\it Prog. Theor. Phys.} {\bf 58}, 816.
\smallskip
Yamamoto, T., and Nishida, S. 1977 {\it Prog. Theor. Phys.} {\bf 57}, 1939.
\vfil
\eject
\centerline{\vbox{\halign{\hfil # \hfil & \hfil # \hfil \cr
\quad Atoms in Cluster \quad&\quad Sticking Coefficient \quad\cr
\noalign{\medskip}
1 & $2.2 \times 10^{-7}$\cr
2 & $3.6 \times 10^{-6}$\cr
3 & $8.2 \times 10^{-5}$\cr
4 & $8.5 \times 10^{-4}$\cr
5 & $5.3 \times 10^{-3}$\cr
6 & $2.4 \times 10^{-2}$\cr
7 & $7.8 \times 10^{-2}$\cr
8 & $0.19$\cr
9 & $0.38$\cr
10 & $0.57$\cr}}}
\bigskip
\centerline{Table 1: Sticking Coefficients}
\vfil
\eject
\vbox{\halign{\hfil # \hfil & \hfil # \hfil & \hfil # \hfil \cr
\quad Atoms in Cluster \quad&\quad Vibrational Degrees of Freedom \quad&
\quad Dissociation Time (sec) \quad\cr
\noalign{\medskip}
3 & 3 & $1.1 \times 10^{-11}$\cr
4 & 6 & $7.2 \times 10^{-9}$\cr
5 & 9 & $1.1 \times 10^{-6}$\cr
6 & 12 & $7.0 \times 10^{-5}$\cr
7 & 15 & $2.6 \times 10^{-3}$\cr
8 & 18 & $6.1 \times 10^{-2}$\cr
9 & 21 & $1.0$\cr
10 & 24 & $14.$\cr}}
\bigskip
\centerline{Table 2: Dissociation Times from RRK Theory for $E = E_d$}
\vfil
\eject
\centerline{\vbox{\halign{\hfil $#$ \hfil & \hfil $#$ \hfil & \hfil $#$ \hfil &
\hfil $#$ \hfil \cr
\quad \lambda T_{1/2} \quad&\quad {\rm Mean\ Energy} \quad&\quad {\rm
Gaussian\ Distribution} \quad&\quad {\rm Exact\ Distribution} \quad\cr
0.167&0.0&1.4 \times 10^7&4.6 \times 10^9\cr
0.333&0.0&5.7 \times 10^7&3.8 \times 10^{10}\cr
0.500&0.0&6.0 \times 10^{10}&1.3 \times 10^{11}\cr
0.667&0.0&2.3 \times 10^{11}&3.2 \times 10^{11}\cr
0.833&0.0&5.4 \times 10^{11}&6.3 \times 10^{11}\cr
1.000&0.0&1.0 \times 10^{12}&1.1 \times 10^{12}\cr
1.167&1.7 \times 10^{11}&1.6 \times 10^{12}&1.6 \times 10^{12}\cr
1.333&7.5 \times 10^{11}&2.4 \times 10^{12}&2.3 \times 10^{12}\cr
1.500&1.7 \times 10^{12}&3.2 \times 10^{12}&3.1 \times 10^{12}\cr
1.667&2.9 \times 10^{12}&4.2 \times 10^{12}&4.0 \times 10^{12}\cr
1.833&4.2 \times 10^{12}&5.2 \times 10^{12}&4.9 \times 10^{12}\cr
2.000&5.5 \times 10^{12}&6.2 \times 10^{12}&5.9 \times 10^{12}\cr
2.167&6.8 \times 10^{12}&7.2 \times 10^{12}&6.9 \times 10^{12}\cr
2.333&8.2 \times 10^{12}&8.3 \times 10^{12}&8.0 \times 10^{12}\cr
2.500&9.5 \times 10^{12}&9.3 \times 10^{12}&9.1 \times 10^{12}\cr
2.667&1.1 \times 10^{13}&1.0 \times 10^{13}&1.0 \times 10^{13}\cr
2.833&1.2 \times 10^{13}&1.1 \times 10^{13}&1.1 \times 10^{13}\cr
3.000&1.3 \times 10^{13}&1.2 \times 10^{13}&1.2 \times 10^{13}\cr
3.167&1.4 \times 10^{13}&1.3 \times 10^{13}&1.3 \times 10^{13}\cr
3.333&1.5 \times 10^{13}&1.4 \times 10^{13}&1.4 \times 10^{13}\cr}}}
\bigskip
\centerline{Table A.1: Trimer Photodissociation Rates (sec$^{-1}$)}
\vfil
\eject
\centerline{\vbox{\halign{\hfil $#$ \hfil & \hfil $#$ \hfil & \hfil $#$ \hfil &
\hfil $#$ \hfil \cr
\quad \lambda T_{1/2} \quad&\quad {\rm Mean\ Energy} \quad&\quad {\rm
Gaussian\ Distribution} \quad&\quad {\rm Exact\ Distribution} \quad\cr
0.167&0.0&1.6 \times 10^{-8}&1.2\cr
0.333&0.0&2.6 \times 10^{-3}&71.\cr
0.500&0.0&0.86&9.2 \times 10^2\cr
0.667&0.0&35.&6.1 \times 10^3\cr
0.833&0.0&5.1 \times 10^2&2.8 \times 10^4\cr
1.000&0.0&4.1 \times 10^3&9.7 \times 10^4\cr
1.167&1.2 \times 10^{-6}&2.2 \times 10^4&2.9 \times 10^5\cr
1.333&1.0 \times 10^{-3}&8.8 \times 10^4&7.3 \times 10^5\cr
1.500&0.11&2.9 \times 10^5&1.7 \times 10^6\cr
1.667&3.6&8.1 \times 10^5&3.6 \times 10^6\cr
1.833&58.&3.0 \times 10^6&7.1 \times 10^6\cr
2.000&5.5 \times 10^2&4.5 \times 10^6&1.3 \times 10^7\cr
2.167&3.7 \times 10^3&9.2 \times 10^6&2.4 \times 10^7\cr
2.333&1.8 \times 10^4&1.8 \times 10^7&4.0 \times 10^7\cr
2.500&7.4 \times 10^4&3.2 \times 10^7&6.5 \times 10^7\cr
2.667&2.5 \times 10^5&5.5 \times 10^7&1.0 \times 10^8\cr
2.833&7.3 \times 10^5&9.1 \times 10^7&1.6 \times 10^8\cr
3.000&1.9 \times 10^6&1.5 \times 10^8&2.3 \times 10^8\cr
3.167&4.4 \times 10^6&2.2 \times 10^8&3.4 \times 10^8\cr
3.333&9.6 \times 10^6&3.3 \times 10^8&4.8 \times 10^8\cr
4.167&1.8 \times 10^8&1.7 \times 10^9&2.1 \times 10^9\cr
5.000&1.3 \times 10^9&6.1 \times 10^9&6.8 \times 10^9\cr}}}
\bigskip
\centerline{Table A.2: 20-mer Photodissociation Rates (sec$^{-1}$)}
\vfil
\eject
\centerline{Figure Captions}
\bigskip
\noindent
Figure 1: Size distribution of grains found without photodissociation.
$M_C$ is the carbon mass and $f$ the number of atoms in a grain.  The units
of the ordinate are arbitrary, but the relative calibration of the two cases
(labeled by the total mass of carbon in the nova debris) is meaningful, as
is the grain radius $a$.
\medskip
\noindent
Figure A.1: Regions of validity (10\% accuracy) of exact, Gaussian, and mean
energy calculations of photodissociation rates.
\vfil
\eject
\end
\bye